\documentclass[twocolumn,showpacs,preprintnumbers,amsmath,amssymb,superscriptaddress,here]{revtex4}
\usepackage{graphicx}
\usepackage{dcolumn}
\usepackage{bm}

\begin{document}

\preprint{APS/123-QED}

\title{
Tensile-Strain Dependent Spin States in Epitaxial LaCoO$_3$ Thin Films
}

\author{Y. Yokoyama}
\affiliation{Institute for Solid State Physics, University of Tokyo, Chiba 277-8581, Japan}
\affiliation{Department of Physics, University of Tokyo, Tokyo 113-0033, Japan}
\author{Y. Yamasaki}
\affiliation{Department of Applied Physics and Quantum-Phase Electronics Center (QPEC), University of Tokyo, Hongo, Tokyo 113-8656, Japan}
\affiliation{RIKEN Center for Emergent Matter Science (CEMS), Wako 351-0198, Japan}
\affiliation{National Institute for Materials Science (NIMS), Tsukuba 305-0047, Japan}
\author{M. Taguchi}
\affiliation{Nara Institute of Science and Technology (NAIST), 89165, Takayama, Ikoma, Nara 630-0192, Japan}
\author{Y. Hirata}
\affiliation{Institute for Solid State Physics, University of Tokyo, Chiba 277-8581, Japan}
\affiliation{Department of Physics, University of Tokyo, Tokyo 113-0033, Japan}
\author{K. Takubo}
\affiliation{Institute for Solid State Physics, University of Tokyo, Chiba 277-8581, Japan}
\author{J. Miyawaki}
\affiliation{Institute for Solid State Physics, University of Tokyo, Chiba 277-8581, Japan}
\author{Y. Harada}
\affiliation{Institute for Solid State Physics, University of Tokyo, Chiba 277-8581, Japan}
\author{D.~Asakura}
\affiliation{Research Institute for Energy Conservation, National institute of Advance Industrial Science and Technology (AIST), Umezono 1-1-1, Tsukuba 305-8568, Japan}
\author{J. Fujioka}
\affiliation{Department of Applied Physics and Quantum-Phase Electronics Center (QPEC), University of Tokyo, Hongo, Tokyo 113-8656, Japan}
\author{M. Nakamura}
\affiliation{RIKEN Center for Emergent Matter Science (CEMS), Wako 351-0198, Japan}
\author{H. Daimon}
\affiliation{Nara Institute of Science and Technology (NAIST), 89165, Takayama, Ikoma, Nara 630-0192, Japan}
\author{M. Kawasaki}
\affiliation{Department of Applied Physics and Quantum-Phase Electronics Center (QPEC), University of Tokyo, Hongo, Tokyo 113-8656, Japan}
\affiliation{RIKEN Center for Emergent Matter Science (CEMS), Wako 351-0198, Japan}
\author{Y. Tokura}
\affiliation{Department of Applied Physics and Quantum-Phase Electronics Center (QPEC), University of Tokyo, Hongo, Tokyo 113-8656, Japan}
\affiliation{RIKEN Center for Emergent Matter Science (CEMS), Wako 351-0198, Japan}
\author{H. Wadati}
\affiliation{Institute for Solid State Physics, University of Tokyo, Chiba 277-8581, Japan}
\affiliation{Department of Physics, University of Tokyo, Tokyo 113-0033, Japan}

\date{\today}

\begin{abstract}
The spin states of Co$^{3+}$ ions in perovskite-type LaCoO$_3$, governed by complex interplay between the electron-lattice interactions and the strong electron correlations,
still remain controversial due to the lack of experimental techniques which can detect directly.
In this letter, we revealed the tensile-strain dependence of spin states, $i. e.$ the ratio of the high- and low-spin states, in epitaxial thin films and a bulk crystal of LaCoO$_3$ via resonant inelastic soft x-ray scattering.
The tensile-strain as small as 1.0\% was found to realize different spin states from that in the bulk.
\end{abstract}

\pacs{78.70.En, 78.70.Dm, 36.40.Mr, 73.43.Fj}
\maketitle
\newpage

Charge, spin, and orbital degrees of freedom activated by strong electron correlation realize
emergent physical phenomena such as superconductivity,
metal-insulator transition, and charge ordering \cite{Imada}.
Perovskite-type cobalt oxide LaCoO$_3$ is one of the most intriguing materials due to its versatile electron degrees of freedom.
Since the spin state of LaCoO$_3$ is sensitive to the crystal field, it shows a spin crossover from the low-spin (LS) to high-spin (HS) state with increasing temperature \cite{Haverkort}.
In addition, the spin states can be modified by external stimuli such as a magnetic field and pressure. For example, the possibility of an excitonic insulating state under a high magnetic field was theoretically proposed in the bulk crystal and has also attracted considerable interest these days \cite{Kunes1,Sotnikov,Nasu}. Epitaxial strain can also influence the spin states as observed for ferromagnetism in LaCoO$_3$ thin films at lower temperatures ($\lesssim$ 85 K) \cite{Fuchs1,Fuchs2,Freeland,Mehta,Rata,Fujioka1,Fujioka2,Yamasaki}.

The crystal structure of bulk LaCoO$_3$ is a rhombohedral perovskite-type
with corner-sharing CoO$_6$ octahedra, in which
the valence of Co is 3+ with 3$d^6$ configuration.
The spin state of LaCoO$_3$ is considered to take
LS state with $e_g^0t_{2g}^6$, intermediate-spin (IS) state with $e_g^1t_{2g}^5$, or HS state with $e_g^2t_{2g}^4$.
The population of HS state gradually increases as the temperature increases from around 100 K and the HS state becomes dominant above 500 K \cite{Haverkort}.
Although the spin state and its temperature dependence of LaCoO$_3$ bulk have already been studied by using various experimental and theoretical approaches \cite{Abbate,Korotin,Magnuson,Podlesnyak,Knizek,Kunes2,Hariki}, the spin states have not been completely determined yet. Especially, the spin states in the intermediate temperature region remain controversial, $i. e.$ the mixed HS/LS state or the IS state.
On the other hand, in ferromagnetic LaCoO$_3$ thin films,
it is indicated that the ferromagnetism originates from the spin-state, orbital, and spin orderings
\cite{Fujioka1,Fujioka2,Yamasaki}.
From the resonant x-ray diffraction, modulation vector $\bm{q}$ = (1/4, $-1/4$, 1/4)$_{pc}$ (the suffix $pc$ stands for the pseudo-cubic setting) and $\bm{q}$ = (1/6, 1/6, 1/6)$_{pc}$ were observed in the thin films grown on LSAT(110) and LSAT(111) substrates, respectively [LSAT: (LaAlO$_3$)$_{0.3}$(SrAl$_{0.5}$Ta$_{0.5}$O$_3$)$_{0.7}$]. Then, the spin states are considered to depend on the tensile-strain. In these thin films, some possible models of spin state orderings were reported in Ref.
\cite{Fujioka1,Fujioka2}.
The proposed relationships between the strain and the estimated ratio of spin states per unit cell are shown in TABLE I.

In previous studies, the spin states were conjectured from the orderings of 3$d$ electrons on the basis of resonant x-ray diffraction \cite{Fujioka1,Fujioka2,Yamasaki}. However, direct observations of the electronic structures are needed to clarify the spin states. Since it is difficult to determine the HS/LS ratio by conventional methods such as x-ray absorption spectroscopy (XAS), we performed resonant inelastic soft x-ray scattering (RIXS) by using Co 2$p$ $\to$ $3d$ $\to$ $2p$ process ($L$-edge). The RIXS is one of the most powerful techniques to clarify the spin states by observing the $d$-$d$ excitations
\cite{Ament,Kotani,Chiuzbaian}.
\begin{table}[b]
\caption{The estimated ratio of spin states per unit cell at the lowest temperature in the previous studies. A positive value of strain indicates a tensile-strain.}
\vskip 3pt
  \begin{tabular}{l@{\hspace{0.7cm}}|@{\hspace{0.2cm}}c@{\hspace{0.2cm}}|@{\hspace{0.2cm}}ccc@{\hspace{0.2cm}}|@{\hspace{0.1cm}}c} \hline 
\hline
     & Strain (\%) & LS : & IS : & HS & Ref. \\ \hline
   LaCoO$_3$/LSAT(110) & 1.0 & 0 & 3 & 1 & \cite{Fujioka1} \\
   LaCoO$_3$/LSAT(111) & 0.5 & 2 & 0 & 1 & \cite{Fujioka2} \\
   LaCoO$_3$ bulk & 0 & 1 & 0 & 0 & \cite{Haverkort} \\ \hline
\hline
  \end{tabular}
\end{table}
A higher-energy-resolution (a few hundreds of meV) is required to distinguish the $d$-$d$ excitations from the elastic scattering \cite{Magnuson}.

\begin{figure}
  \begin{center}
    \includegraphics[width=0.9\linewidth]{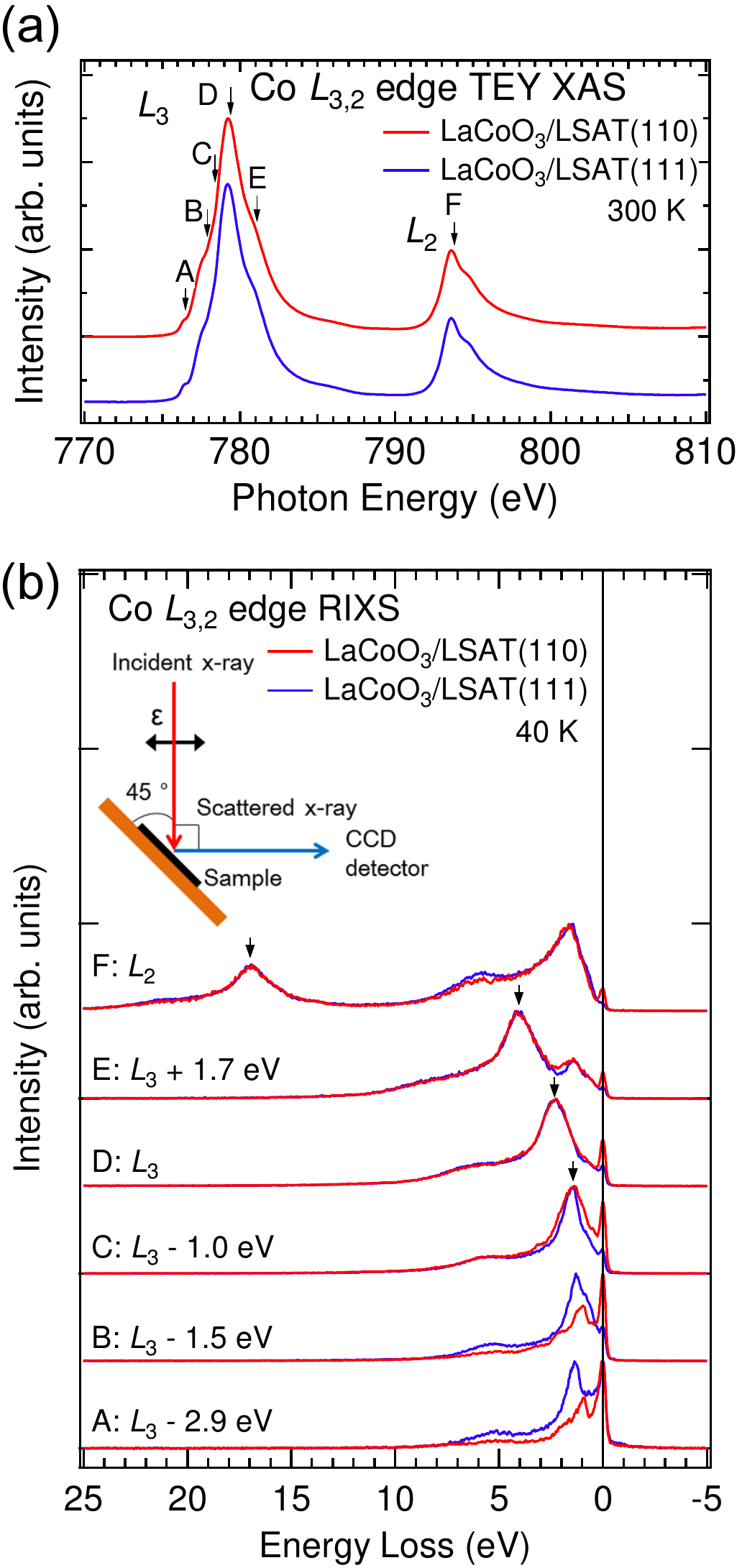}
  \end{center}
  \caption{(Color online) Spectra measured at Co $L_{3,2}$ edge by (a) TEY-XAS and (b) RIXS spectra of LaCoO$_3$ thin films grown on LSAT(110) and LSAT(111) substrates. The arrows and letters in XAS indicate the excitation energies for RIXS. All RIXS spectra are normalized by the intensity of the highest peak and the arrows indicate the fluorescence peaks. The inset shows the schematic diagram of the experimental setup in our RIXS measurements.
}
\end{figure}

The LaCoO$_3$ epitaxial thin films with the thickness of 30 nm were fabricated on LSAT(110) and LSAT(111) substrates by pulsed laser deposition technique. Details for the syntheses and the characterizations were described on Ref. \cite{Fujioka1,Fujioka2}. The lattice constant of LSAT substrate is 3.868 \AA, while that of LaCoO$_3$ bulk is
3.804 \AA \cite{Kobayashi}, meaning that the LaCoO$_3$ epitaxial thin films grown on LSAT substrates are distorted by tensile-strain.
The tensile-strains from LSAT(110) and LSAT(111) substrates are 1.0\%
and 0.5\%, respectively (strains are defined as the ratio of the cubic root of the unit cell volume) \cite{Fujioka1,Fujioka2}.
The experiments of XAS and RIXS were performed at BL07LSU HORNET, SPring-8 \cite{Harada}.
Before RIXS measurements, we obtained the XAS spectra at 300 K by total electron yield (TEY) in order to determine the excitation energy of RIXS.
The RIXS measurements were performed at 40 K and 300 K by using soft x-ray from 770 eV to 810 eV (Co $L_{3,2}$ edge). In the range of the x-ray energy, energy resolution $\Delta E$ $\sim$ 300 meV, which was determined via full width of half maximum (FWHM) of the elastic peak.  
The charge coupled device (CCD) detector was set at 90$^\circ$ relative to the incident x-ray with horizontal polarization to suppress the elastic scattering (see inset in Fig. 1(b)).

\begin{figure}[t]
  \begin{center}
\includegraphics[width=0.9\linewidth]{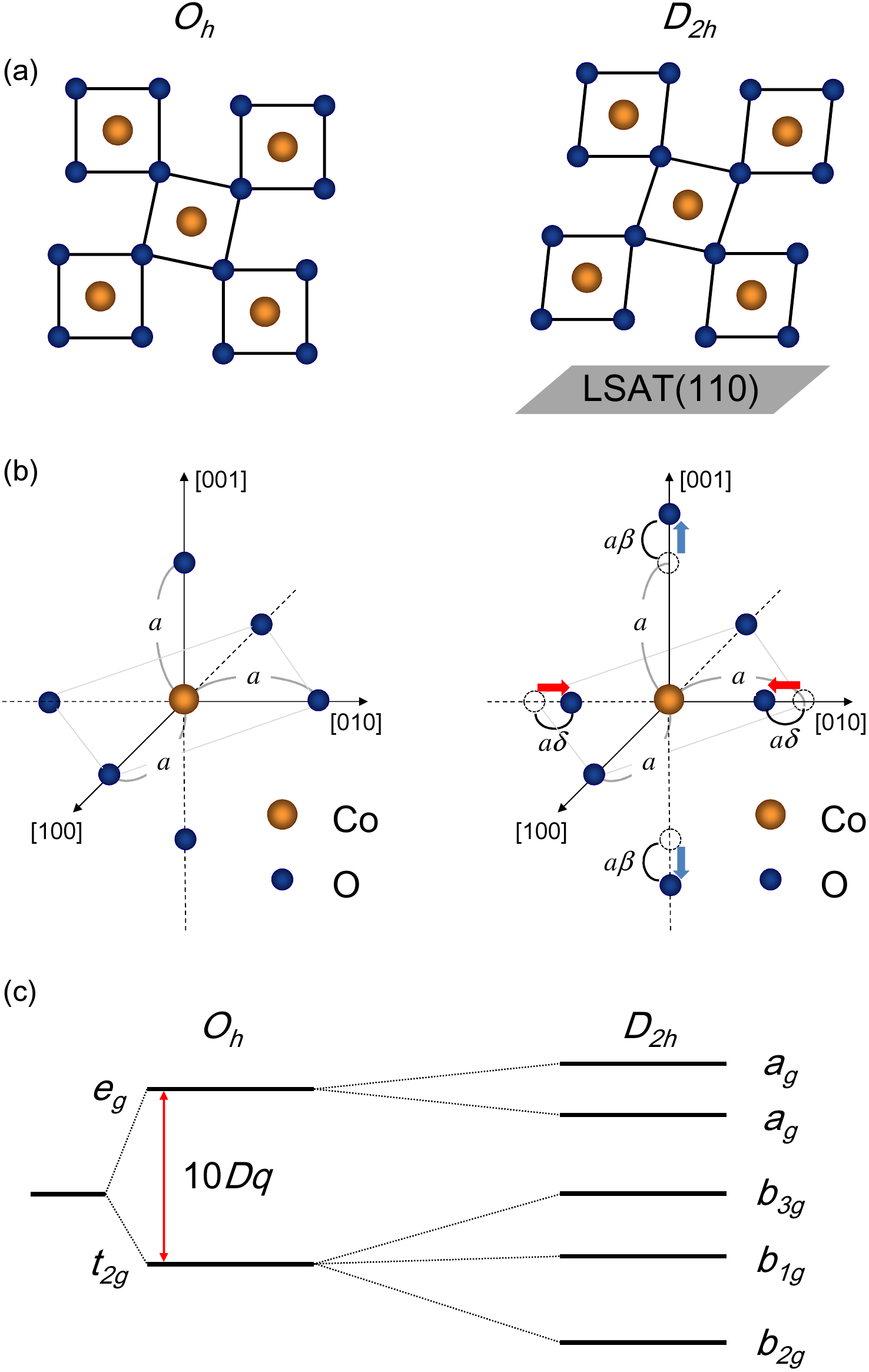}
  \end{center}
  \caption{(Color online) Schematic view of (a) the crystal structure and (b) [CoO$_6$]$^{9-}$ cluster in $O_h$ and $D_{2h}$ symmetry. ``$a$'' correspond to the bond distance between Co and O ions in LaCoO$_3$ bulk, while ``$\delta$'' and ``$\beta$'' were defined as the scaling factors of the crystal distortion. (c) The energy level diagram of $O_h$ and $D_{2h}$ symmetry in our calculations. 10$Dq$ indicates the crystal field splitting in $O_h$ symmetry.
}
\end{figure}

First, we measured Co $L_{3,2}$ XAS by TEY mode of LaCoO$_3$ thin films at 300 K. The spectra are shown in Fig. 1(a). The peaks around 779 eV and 794 eV correspond to the Co $L_3$ and $L_2$ edges, respectively. Although the magnitude of the tensile-strain and modulation vector $q$ are different between LSAT(110) and LSAT(111), the spectral shapes are quite similar.
To investigate the detailed electronic structures from $d$-$d$ excitations, we performed RIXS measurements.
As the excitation energy of RIXS, we selected the energy of A: $L_3-2.9$ eV, B: $L_3-1.5$  eV, C: $L_3-1.0$ eV, D: $L_3$, E: $L_3+1.7$ eV, and F: $L_2$.
The Co $L_{3,2}$ RIXS spectra of the LaCoO$_3$ thin films are shown in Fig. 1(b).
These spectra are normalized by the intensity of the highest peak.
In this figure, the thick solid line shows the center of elastic scattering peaks and the arrows
indicate the fluorescence ones. Other peaks from 0 eV to 4 eV correspond to the $d$-$d$ excitations and the peak around 5.0 eV may be a charge-transfer (CT) excitation.
In the spectra of A and B, the peaks of $d$-$d$ excitations are clearly observed.
We find that the $d$-$d$ excitations are obviously different between
the thin films on LSAT(110) with 1.0\% tensile-strain and LSAT(111) with 0.5\% tensile-strain; this indicates that the spin states of LaCoO$_3$ change drastically according to the magnitude of the tensile-strain.
On the other hand, in the spectra excited by higher energy (C, D, E, and F),
the peaks of $d$-$d$ excitations are not clear because of the larger fluorescence.
Then, we selected the excitation energy A: $L_3-2.9$ eV and compare with the
RIXS spectra of bulk single crystal in order to discuss the relationship between the spin states and the epitaxial strain.

\newlength{\myheight}
\setlength{\myheight}{1cm}
\begin{table}[b]
\caption{Parameter values used in our calculations based on the impurity Anderson model. For details of these parameters, see main text. Units are in eV.}
\vskip 3pt
  \begin{tabular}{l@{\hspace{0.7cm}}c@{\hspace{0.5cm}}c@{\hspace{0.5cm}}c@{\hspace{0.5cm}}c@{\hspace{0.5cm}}c@{\hspace{0.5cm}}c} \hline 
\hline
    & $\Delta$ & $\Delta^*$ & $10Dq$ & $(pd\sigma)$ & $\delta$ & $\beta$ \\ \hline
   HS($O_h$) & 1.1 & 0.4 & 1.38 & $-1.96$ & 0 & 0 \\
   LS($O_h$) & 1.8 & 0.7 & 1.40 & $-2.00$ & 0 & 0 \\ 
   HS($D_{2h}$) & 0.2 & 0.1 & 1.40 & $-2.00$ & 0.010 & 0.005 \\
\hline
\hline
  \end{tabular}
\end{table}

\begin{figure}[t]
  \begin{center}
    \includegraphics[width=0.9\linewidth]{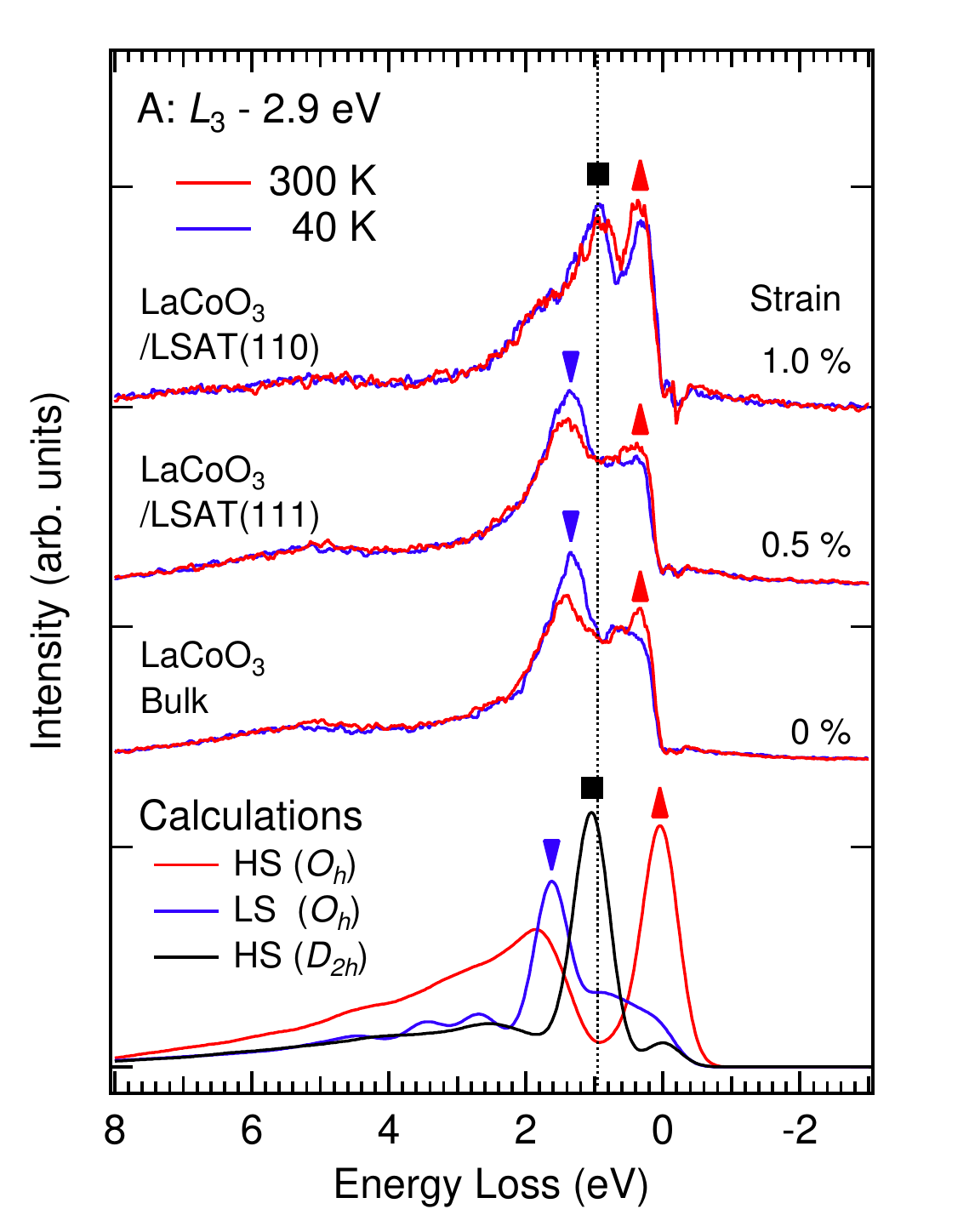}
  \end{center}
  \caption{(Color online) Temperature dependence (300 K \& 40 K) of the RIXS spectra
excited at A: $L_3-2.9$ eV (776.5 eV) in comparison with the impurity Anderson model calculations in $O_h$ and $D_{2h}$ symmetries.
The experimental spectra were normalized by their area after subtracting the elastic scatterings by Gaussian fitting.
The triangles, squares, and inverted triangles indicate the main peak of the HS($O_h$), HS($D_{2h}$), and LS($O_h$) states, respectively.
}
\end{figure}

In the analysis, we carried out impurity Anderson model calculations in $O_h$ and $D_{2h}$ local symmetry.
Figure 2(a) show the schematic figures of the crystal structures in bulk and thin film on LSAT(110).
The symmetry of bulk is nearly $O_h$, while that of the thin film on LSAT(110) and LSAT(111) are $D_{2h}$ and $D_{3d}$, respectively.
In the film on LSAT(111), the strain is so small that we consider its symmetry as $O_h$.
Therefore, we performed the calculations in $O_h$ symmetry for the LaCoO$_3$ bulk and the thin film on LSAT(111), while in $D_{2h}$ symmetry for the thin film on LSAT(110).
As shown in Fig. 2(b), we consider the [CoO$_6$]$^{9-}$ cluster model with the scaling factors $\delta$ and $\beta$ in order to describe the crystal distortion, $i.e.$ shrinkage along [010] axis and elongation along [001] axis.
In the present model, we consider the Co $3d$ orbitals of a central Co atom (core hole site) and appropriate linear combinations of Co $3d$ orbitals on neighboring sites. 
The anisotropic effect due to the lattice mismatch of substrate is also taken into account through the anisotropic hybridization and the crystal field for Co $3d$ states, where the local symmetry around the Co ion is approximately treated as $D_{2h}$. As shown in Fig. 2(c), the energy diagram in $D_{2h}$ symmetry is different from that in $O_h$ symmetry, where 10$Dq$ is defined as the crystal field splitting between $e_g$ and $t_{2g}$ orbitals. In $D_{2h}$ symmetry, the two energy levels ($e_g$ and $t_{2g}$) are split into five energy levels by reducing the symmetry, indicating that the new spin states can be realized.
In our calculations, the charge-transfer energy from the O $2p$ ligand band to the upper Hubbard band and that from the non-local band to the upper Hubbard band are defined as 
 $\Delta$ and $\Delta^*$, respectively. The Coulomb interaction between Co $3d$ states ($U_{dd}$) and that between Co $3d$ and $2p$ core-hole states ($-U_{dc}$) were set as $U_{dd}=5.6$ eV and $U_{dc}=7.0$ eV, respectively. The meaning of these parameters were defined as in Ref. \cite{Taguchi}. The Slater integrals and the spin-orbit coupling constant are calculated by Cowan's Hartree-Fock program \cite{Cowan} and then the Slater integrals are rescaled by 85$\%$, as usual.
($pd\sigma$) and ($pd\pi$) are the transfer integrals between Co $3d$ and O $2p$ orbitals and we used the parametrization given in Harrison's rule as $(pd\sigma) \propto d^{-3.5}$ \cite{Harrison}. ($d$ denotes the atomic distance.)
In our calculations, the parameters were set as TABLE II. The spin crossover (HS $\leftrightarrow$ LS) occur between 10$Dq = 1.38$ and 1.40 eV in the $O_h$ symmetry. The value of $\delta$ and $\beta$ in the case of the $D_{2h}$ symmetry were obtained from lattice parameters \cite{Fujioka1}. Note that the calculations were performed in the wide range of parameters (10$Dq$: 0 - 4.00 eV, $\beta$: 0 - 0.040, $\delta$: 0 - 0.040, and ($pd\sigma$): $-$3.00 - $-$1.00 eV) and there is no parameter region where the IS state is the most stable.

\begin{figure}[t]
  \begin{center}
    \includegraphics[width=\linewidth]{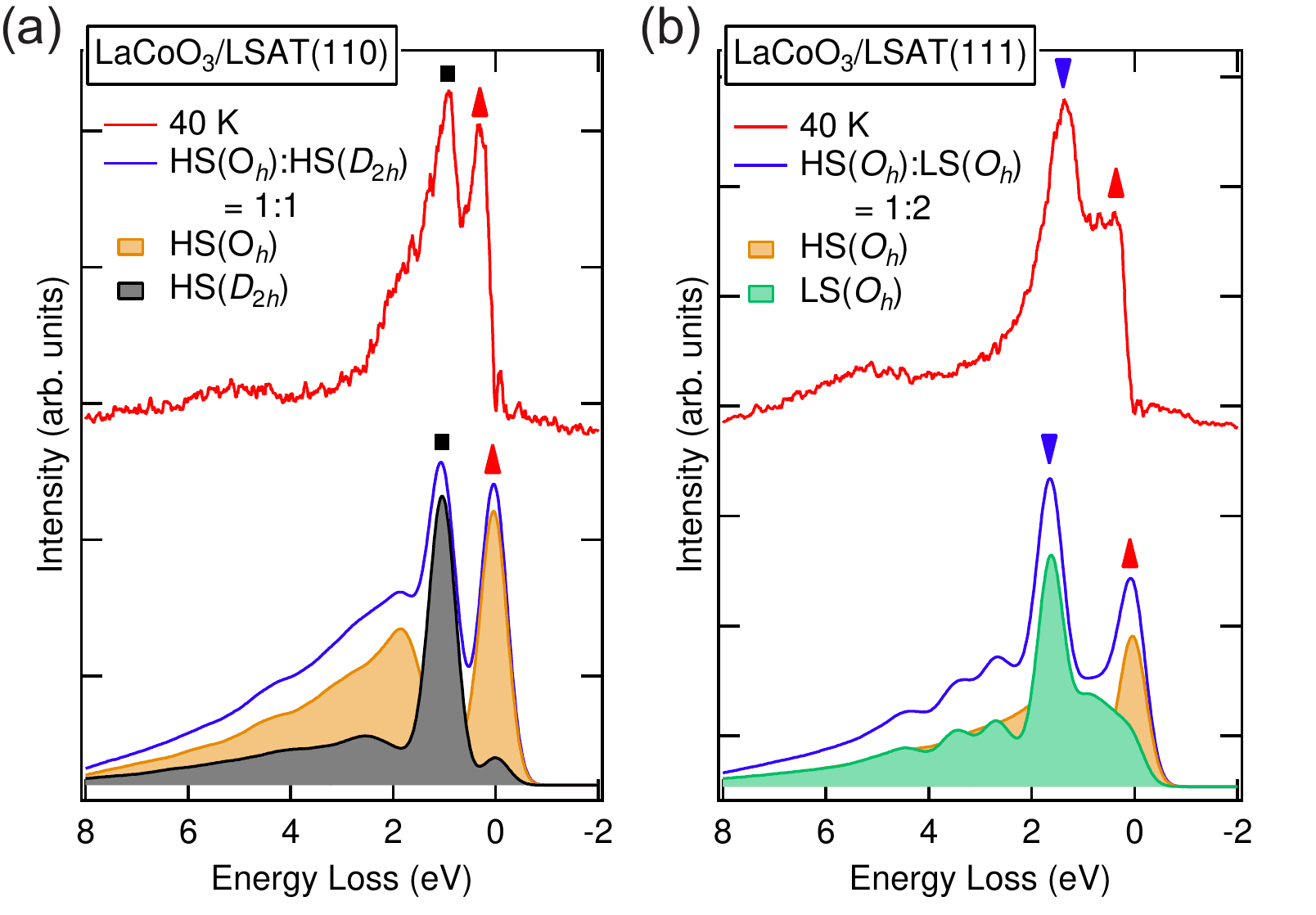}
  \end{center}
  \caption{(Color online) Comparison between the experimental RIXS and the linear combinations of theoretical spectra in (a) LaCoO$_3$/LSAT(110) and (b) LaCoO$_3$/LSAT(111). The triangles, squares, and inverted triangles indicate the main peak of the HS($O_h$), HS($D_{2h}$), and LS($O_h$) grand states, respectively.
}
\end{figure}

The experimental RIXS spectra excited with A: $L_3-2.9$ eV (776.5 eV) measured at 40 K and 300 K are shown in comparison to the theoretical ones with several electronic states in Fig. 3.
There are two kinds of strong peaks in the energy range between 0 and 2 eV in all three samples. 
The peaks observed around 0.3 eV can be assigned to the excitation originated form the HS states.
On the other hand, the peaks around 1.3 eV observed in LaCoO$_3$/LSAT(111) and bulk correspond to the LS ground states.
As the temperature increasing, the behavior that the peak intensities at 0.3 eV increase and those at 1.3 eV decrease is consistent with the fact that population of the HS state increases with increasing temperature.
However, peaks which appear at 1.0 eV in LaCoO$_3$/LSAT(110) can be explained by neither the LS nor HS with $O_h$ symmetry.
As seen in the theoretical spectra, the peak for HS state is shifted to 1.0 eV by lowering the symmetry from $O_h$ to $D_{2h}$. 
It indicates that the spin state of LaCoO$_3$/LSAT(110) consists of the HS states with different local symmetries, $i.e.$ the mixture of $O_h$ and $D_{2h}$ symmetries.   

To compare the present results with the previous study by resonant x-ray diffraction, we estimate the ratio of spin states by the theoretical spectra of various spin states.
Since the theoretical spectra are normalized by the intensity of the theoretical XAS at A: $L_3-2.9$ eV, we can reproduce the mixed spin states by linear combinations of the spectra and estimate the ratio of the spin states by the peak intensity in each spin state.
As shown in Fig. 4(a), HS($O_h$) : HS($D_{2h}$) = 1 : 1 is in good agreement with the experimental spectrum of LaCoO$_3$/LSAT(110), indicating that the CoO$_6$ octahedra in $O_h$ and $D_{2h}$ local symmetry coexist as the ratio of 1 : 1.
Since the spatial modulation of spin state is fourfold periodicity, one plausible model is that the spin states are aligned in order of $O_h$ - $D_{2h}(d_{yz})$ - $O_{h}$ - $D_{2h}(d_{zx})$, where the HS states with $D_{2h}$ have two different orbital states.
In this study, we did not find the theoretical spectra of IS states which reproduce experimental results and thus could not confirm the spin state model of IS : HS = 3 : 1.

In conclusion, the spin states of LaCoO$_3$ change from LS to HS according to the magnitude of the epitaxial strain, which could not be detected by conventional XAS but could be clarified by RIXS measurements.
The LS ground state in bulk fully changes to HS states by tensile-strain as small as 1.0\%.
Especially in the thin film on LSAT(110), we observed the tensile-strain induced unique spin state, $i. e.$ HS($O_h$) : HS($D_{2h}$) = 1 : 1. Since the spin state is not stabilized by temperature or hydrostatic pressure in bulk, applying the tensile-strain is an important factor for realizing novel spin states.

The authors thank T. Arima and K. Yamamoto for productive discussions.
The present work was performed under the approvals of SPring-8 Japan Synchrotron Radiation Research Institute (Proposal No. 2016A7501, No. 2016B7515).

\end{document}